\documentclass[floatfix,aps,prl,twocolumn,showpacs,showkeys,10pt]{revtex4}
\usepackage{epsfig}
\usepackage{epsf}
\usepackage{amsmath}
\usepackage{amssymb}
\begin{document}


\title{Chromo-polarizability and $\pi\pi$ final state interaction}

\author{Feng-Kun Guo,$^{1,2,6}$ Peng-Nian Shen,$^{2,1,4,5}$}
\author{Huan-Ching Chiang$^{3,1,5}$}
\affiliation{$^1$Institute of High Energy Physics, Chinese Academy of Sciences,
P.O.Box 918(4), Beijing 100049, China \\
$^2$CCAST(World Lab.), P.O.Box 8730, Beijing 100080, China\\
$^3$South-West University, Chongqing 400715, China\\
$^4$Institute of Theoretical Physics, Chinese Academy of Sciences, P.O.Box 2735, China\\
$^5$Center of Theoretical Nuclear Physics, National Laboratory of
Heavy Ion Accelerator, Lanzhou 730000, China\\
$^6$Graduate University of Chinese Academy of Sciences, Beijing
100049, China}

\date{\today}

\begin{abstract}
The chromo-polarizability of a quarkonium state is a measure of the
amplitude of the $E1$-$E1$ chromo-electric interaction of the
quarkonium with soft gluon fields and can be measured in the heavy
quarkonium decays. Based on the chiral unitary approach, formulas
with modification caused by the $S$ wave $\pi\pi$ final state
interaction (FSI) for measuring the chromo-polarizabilities are
given. It is shown that the effect of the $S$ wave $\pi\pi$ FSI is
very important in extracting chromo-polarizabilities from the
experimental data. The resultant values with the FSI are reduced to
about 1/3 of those determined without the FSI. The consequences of
the FSI correction in the $J/\psi$-nucleon scattering near the
threshold are also discussed. The estimated lower bound of the total
cross section is reduced from about 17 mb to 2.9 mb, which agrees
with the experimental data point and is compatible with the
previously estimated values in the literature. In order to
understand the interaction of heavy quarkonia with light hadrons at
low energies better and to obtain the chromo-polarizabilities of
quarkonia accurately, more data should be accumulated. This can be
done in the $J/\psi \to \pi^+\pi^-l^+l^-$ decay at BES-III and
CLEO-c and in the $\Upsilon \to \pi^+\pi^-l^+l^-$ decay at B
factories.
\end{abstract}



\maketitle

It was proposed that the suppression of the $J/\psi$ production
could be regarded as a signature for the formation of a quark-gluon
plasma (QGP) due to the color screening \cite{ms86}. The interaction
of $J/\psi$ with light hadrons is very important in studying the
$J/\psi$ suppression in heavy ion collision and in obtaining an
unambiguous signal of the QGP \cite{ba03}. Many studies on this
subject have been done by using various methods and broad ranges of
theoretical predictions have been resulted, for a review, see Ref.
\cite{ba03}.

Recently, Sibirtsev and Voloshin studied \cite{sv05} the interaction
of $J/\psi$ and $\psi'$ with nucleons at low energies near threshold
by employing the multipole expansion method \cite{go78,vo79} and
low-energy theorems in QCD \cite{vz80}. They argued that the total
cross section of the $J/\psi$-nucleon elastic scattering at the
threshold is very likely to be larger than 17 mb which is
significantly larger than the previously estimated values in the
literature. Their result supports the existence of possible bound
light $J/\psi$-nucleus states. Comparing with the experimental value
of $\sigma_{J/\psi N}=3.8\pm0.8\pm0.5$ mb at $\sqrt{s}\approx 5.7$
GeV \cite{ex77}, the large cross section value given in Ref.
\cite{sv05} suggested a noticeable rise towards the threshold. It
should be mentioned that such a result is based on the
chromo-polarizability $\alpha_{J/\psi}$ which appears in describing
the interaction of $J/\psi$ with soft gluons. However, this value is
unknown so far. Although a value of 2 GeV$^{-3}$ of
$|\alpha_{\psi'J/\psi}|$ was used as a reference, it was argued that
the actual value could be somewhat larger. The value of
$|\alpha_{\psi'J/\psi}|$ was estimated by Voloshin via the $\psi'\to
J/\psi\pi^+\pi^-$ decay process by using the QCD multipole expansion
method \cite{vo04}.

The low energy interaction of a heavy quarkonium with a light hadron
is mediated by soft gluons. The heavy quarkonium has a small size
comparing with the wavelengths of gluons, so that one can use a
multipole expansion to study the heavy quarkonium interaction with a
soft gluon \cite{go78,vo79}. The leading $E1$ chromo-electric dipole
term in the expansion is simply
$H_{E1}=-\frac{1}{2}\xi^a\vec{r}\cdot\vec{E}^a(0)$ \cite{go78,vo79},
where $\xi^a$ is the difference between the color generators acting
on the quark and anti-quark, respectively, and can be expressed as
$\xi^a=t_1^a-t_2^a$ with $t_i^a=\lambda_i^a/2$ and $\lambda_i^a$
being Gell-Mann matrices. $\vec{r}$ is the coordinate of the
relative position between the quark and anti-quark. The QCD coupling
constant $g$ is included in the definition of the normalized gluon
field. Using the notations in Ref. \cite{sv05}, the lowest order
amplitude of the transition between two $^3S_1$ heavy quarkonium
states $A$ and $B$ with an emitted gluon system which possesses the
quantum numbers of the dipion is in the second order of $H_{E1}$
\begin{equation}
\langle B|H_{eff}|A\rangle=-\frac{1}{2}\alpha_{AB} \vec{E}^a\cdot
\vec{E}^a,
\end{equation}
where the nonrelativistic normalization is used for the quarkonium
states. $\alpha_{AB}$ is the chromo-polarizability which is a
measure of the amplitude of the $E1$-$E1$ chromo-electric
interaction of the quarkonium with soft gluon fields and is related
to the Green's function $G_A$ for a heavy quark pair in the
color-octet state.

The amplitude for the $\pi^+\pi^-$ transition between $A$ and $B$
can be written as
\begin{equation}
T_{AB}=2\sqrt{M_AM_B}\alpha_{AB}\langle
\pi^+\pi^-|\frac{1}{2}\vec{E}^a\cdot \vec{E}^a|0\rangle,
\end{equation}
where the factor $2\sqrt{M_AM_B}$ appears due to the relativistic
normalization of the decay amplitude $T_{AB}$. It was pointed out
that the dominant part of the two-pion production amplitude by the
gluonic operator $\vec{E}^a\cdot \vec{E}^a$ can be determined by the
trace anomaly and chiral algebra \cite{vz80}
\begin{eqnarray}
\label{eq:ee} \langle \pi^+\pi^-|\vec{E}^a\cdot \vec{E}^a|0\rangle
&=& \frac{8\pi^2}{b}q^2 + O(\alpha_sq_0^2) + O(m_{\pi}^2)
\nonumber\\
&\approx& \frac{8\pi^2}{b}(q^2-C),
\end{eqnarray}
where $q$ is the total four-momentum of the produced dipion system,
$q_0$ is the total energy of the system, $b=9$ is the first
coefficient in the QCD $\beta$ function with three light flavors,
and $C$ is a constant to evaluate approximately the contributions of
sub-leading terms \cite{vz87,vo04} which vary relatively slow with
$q^2$ in the physical region of the pionic transitions \cite{ns81}.
Theoretically, the chromo-polarizability is, at least, highly
model-dependent \cite{vo04} in the present stage. Therefore, it
would be nice if the values of $\alpha_{AB}$ could be extracted from
experimental data. Voloshin performed such an analysis \cite{vo04}
and gave the estimations of $|\alpha_{\psi'J/\psi}|\approx 2.0$
GeV$^{-3}$ and $|\alpha_{\Upsilon'\Upsilon}|\approx 0.66$
GeV$^{-3}$. In the same reference, it was also suggested that the
chromo-polarizabilities of $J/\psi$ and $\Upsilon$ can directly be
measured in the decays $J/\psi\to\pi^+\pi^-l^+l^-$ and
$\Upsilon\to\pi^+\pi^-l^+l^-$ with soft pions. Based on an
approximation that the intermediate state to the lepton pairs is the
$1^3S_1$ quarkonium state itself, the diagonal polarizabilities
(i.e. $\alpha_{AA}$, written as $\alpha_A$ for simplicity) of
$J/\psi$ and $\Upsilon$ can be written as \cite{vo04}
\begin{eqnarray}
\label{eq:wll}
d\Gamma(1^3S_1\to\pi^+\pi^-l^+l^-)&\!\!=&\!\!\frac{(q^2-C)^2}{4b^2q_0^2}
|\alpha_{1S}|^2\sqrt{1-\frac{4m_{\pi}^2}{q^2}}
\nonumber\\
&& \times\sqrt{q_0^2-q^2} \Gamma_{ee}(1^3S_1)dq^2dq_0,\nonumber\\
&&
\end{eqnarray}
where $\Gamma_{ee}(1^3S_1)\equiv \Gamma(1^3S_1\to l^+l^-)$ is the
leptonic width of the $1^3S_1$ state.

On the other hand, the $S$ wave $\pi\pi$ FSI plays an important role
in the heavy quarkonium $\pi^+\pi^-$ transitions \cite{lr02,gs05}.
This FSI would modify the $\pi^+\pi^-$ production amplitude and
consequently would change the determined chromo-polarizabilities.
There are various ways to account for the $S$ wave $\pi\pi$ FSI. One
of the efficient methods is the coupled-channel chiral unitary
approach (ChUA) \cite{oo97}.

By employing such an approach, the $S$ wave $\pi\pi$ phase shift
data can be well described with one parameter, a 3-momentum cut-off
$q_{max}=1.03$ GeV, only, and the low-lying scalar mesons ($\sigma$,
$f_0(980)$, $a_0(980)$ and $\kappa$) with reasonable masses and
widths can dynamically be generated \cite{oo97,gp05}. For the $S$
wave isoscalar channel, the $\pi\pi$ and $K{\bar K}$ channels are
taken into account (for detailed information, refer to Ref.
\cite{oo97}). The normalizations used are
$\langle\pi^+\pi^-|\pi^+\pi^-\rangle =
\langle\pi^-\pi^+|\pi^-\pi^+\rangle =
\langle\pi^0\pi^0|\pi^0\pi^0\rangle = 2$ and $|\pi\pi\rangle^{I=0} =
(|\pi^+\pi^-+\pi^-\pi^++\pi^0\pi^0\rangle)/\sqrt{6}$. Thus the full
amplitude of the process
$\pi^+\pi^-+\pi^-\pi^++\pi^0\pi^0\to\pi^+\pi^-$ can be expressed as
\begin{eqnarray}
\langle\pi^+\pi^-|T|\pi^+\pi^-+\pi^-\pi^++\pi^0\pi^0\rangle =
2T^{I=0}_{\pi\pi,\pi\pi},
\end{eqnarray}
where $T^{I=0}_{\pi\pi,\pi\pi}$ is the full $S$ wave $\pi\pi\to
\pi\pi$ coupled-channel amplitude for $I=0$. Note that the $K{\bar
K}$ channel appears in the intermediate states \cite{oo97}.

By considering the $S$ wave $\pi\pi$ FSI by ChUA
\cite{oo98,mo01,gs05}, the decay amplitude of the $A\to B\pi^+\pi^-$
process is modified to
\begin{equation}
\label{eq:tab}
T_{AB}=\frac{8\pi^2}{b}\sqrt{M_AM_B}\alpha_{AB}(q^2-C)
(1+2G_{\pi}(q^2)T^{I=0}_{\pi\pi,\pi\pi}(q^2)),
\end{equation}
where $G_{\pi}(q^2)$ is the two-pion loop integral
\begin{equation}
\label{eq:2loop}
G_{\pi}(q^2)=i\int\frac{d^4q}{(2\pi)^4}\frac{1}{q^2-m_{\pi}^2+i
\varepsilon} \frac{1}{(p^{^{\prime}}-p-q)^2-m_{\pi}^2+i\varepsilon},
\end{equation}
where $p'$ and $p$ represent the momenta of the initial and the
final quarkonium states, respectively. The loop integration can be
calculated with the same cut-off momentum, $q_{max}=1.03$ GeV, as
the one used in describing the $S$ wave $\pi\pi$ scattering data
\cite{oo97}. In this way, the consistency between the used $\pi\pi$
FSI and the $\pi\pi$ scattering is guaranteed. Note that the direct
$\pi\pi$ production amplitude in Eq. (\ref{eq:ee}) is independent of
the momentum carried by one of the produced pion, i.e. the
integrated loop momentum, hence can be factorized out from the loop
integral. According to Refs. \cite{oo97,oo99,na00}, the amplitude
$T^{I=0}_{\pi\pi,\pi\pi}$ in Eq. (\ref{eq:tab}) can be factorized
out from the loop integral.

\begin{figure}[htb]
\begin{center}\vspace*{2.cm}
{\epsfysize=6cm \epsffile{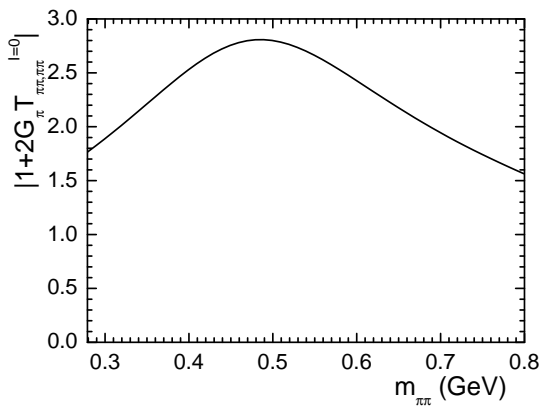}}%
\vglue -2.8cm\caption{\label{fig1}
$|1+2G_{\pi}(q^2)T^{I=0}_{\pi\pi,\pi\pi}(q^2)|$ as a function of the
invariant mass of $\pi^+\pi^-$.}
\end{center}
\end{figure}

Now, in terms of Eq. (\ref{eq:tab}), one can easily work out the
differential width for the $A\to B\pi^+\pi^-$ decay
\begin{widetext}
\begin{equation}
\label{eq:dw}%
d\Gamma(A\to B\pi^+\pi^-) = \frac{2\pi M_B}{b^2M_A}
|\alpha_{AB}|^2(q^2-C)^2
|1+2G_{\pi}(q^2)T^{I=0}_{\pi\pi,\pi\pi}(q^2)|^2
\sqrt{\frac{q^2}{4}-m_{\pi}^2}q_Bdm_{\pi\pi},
\end{equation}
\end{widetext}
where the polarizations of particle $A$ are averaged, and the
polarizations of particle $B$ are summed, and $q_B$ is the magnitude
of the 3-momentum of the $B$ state in the rest frame of particle $A$
\begin{equation}
q_B=\frac{1}{2M_A}\sqrt{(M_A^2-(m_{\pi\pi}+M_B)^2)(M_A^2-(m_{\pi\pi}-M_B)^2)}.
\end{equation}
To demonstrate the effect of the $\pi\pi$ FSI on the differential
width, we plot the factor
$|1+2G_{\pi}(q^2)T^{I=0}_{\pi\pi,\pi\pi}(q^2)|$ versus the invariant
mass of $\pi^+\pi^-$, $m_{\pi\pi}=\sqrt{q^2}$, in Fig. \ref{fig1}.
From the figure, one sees that around $m_{\pi\pi}=0.48$ GeV, this
factor could be enhanced up to $2.8$ from unity by the FSI.

\begin{figure}[htb]
\begin{center}\vspace*{2.cm}
{\epsfysize=6cm \epsffile{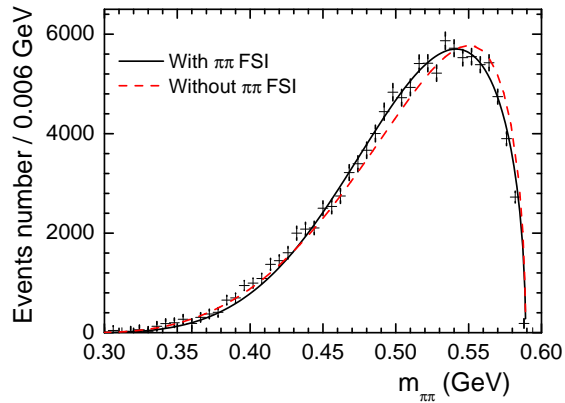}}%
\vglue -2.8cm\caption{\label{fig2}The $\pi^+\pi^-$ invariant mass
spectrum for the $\psi'\to J/\psi\pi^+\pi^-$ decay. The solid and
dashed curves are calculated in the with $\pi\pi$ FSI and without
$\pi\pi$ FSI cases, respectively.}
\end{center}
\end{figure}

\begin{figure}[htb]
\begin{center}\vspace*{2.cm}
{\epsfysize=6cm \epsffile{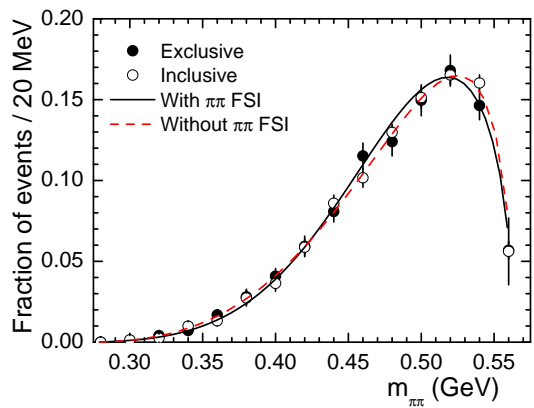}}%
\vglue -2.8cm\caption{\label{fig3}The $\pi^+\pi^-$ invariant mass
spectrum for the decay $\Upsilon'\to\Upsilon\pi^+\pi^-$. The solid
and dashed curves are calculated in the with $\pi\pi$ FSI and
without $\pi\pi$ FSI cases, respectively.}
\end{center}
\end{figure}

In order to determine the transition chromo-polarizabilities
$\alpha_{\psi'J/\psi}$ and $\alpha_{\Upsilon'\Upsilon}$, we fit the
theoretical results to the experimental $\pi^+\pi^-$ invariant mass
spectra from the BES $\psi'\to J/\psi\pi^+\pi^-$ decay data
\cite{bes} and the CLEO $\Upsilon'\to \Upsilon\pi^+\pi^-$ decay data
\cite{cleo}, and the corresponding decay widths $\Gamma(\psi'\to
J/\psi\pi^+\pi^-)=715.8$ keV and
$\Gamma(\Upsilon'\to\Upsilon\pi^+\pi^-)=8.08$ keV \cite{pdg} by
adjusting C and $\alpha_{AB}$ in Eq. (\ref{eq:dw}) as free
parameters. The fitted $\pi^+\pi^-$ invariant spectra for $\psi'\to
J/\psi\pi^+\pi^-$ and $\Upsilon'\to \Upsilon\pi^+\pi^-$ decays are
plotted in Fig. \ref{fig2} and Fig. \ref{fig3}, respectively. In
these figures, the solid and dashed curves represent the results
with and without the $S$ wave $\pi\pi$ FSI, respectively. The
resultant parameters through the best fitting are listed in Table
\ref{tab:para}. It is shown that the $S$ wave $\pi\pi$ FSI provides
rather large modifications to the values of $|\alpha_{\psi'J/\psi}|$
and $|\alpha_{\Upsilon'\Upsilon}|$. The resultant values with the
$\pi\pi$ FSI are almost 1/3 of the those without the $\pi\pi$ FSI.
It should be emphasized that such a modification is meaningful,
because no more free parameters are adopted when the $\pi\pi$ FSI is
included. It should also be mentioned that the parameter values
obtained in the without FSI case are slightly different with those
given by Voloshin, because some approximations was made in the width
formula in Ref. \cite{vo04}.
\begin{table}[htb]
\caption{\label{tab:para} Parameters used in the heavy quarkonium
$\pi^+\pi^-$ transitions with and without the $\pi\pi$ FSI.}
\begin{ruledtabular}
\begin{tabular}{lcc}
   & Without $\pi\pi$ FSI &  With $\pi\pi$ FSI \\ \hline%
$|\alpha_{\psi'J/\psi}|$ (GeV$^{-3}$) & $2.24\pm0.02$ & $0.83\pm0.01$ \\%
$|\alpha_{\Upsilon'\Upsilon}|$ (GeV$^{-3}$) & $0.70\pm0.01$ & $0.24\pm0.01$ \\%
$C_{\psi'\to J/\psi\pi^+\pi^-}$ ($m_{\pi}^2$) &  $4.27\pm0.07$  & $4.25\pm0.09$ \\%
$C_{\Upsilon'\to \Upsilon\pi^+\pi^-}$ ($m_{\pi}^2$) &  $3.45\pm0.15$  & $2.86\pm0.21$ \\%
\end{tabular}
\end{ruledtabular}
\end{table}

In the same way, Eq. (\ref{eq:wll}) for determining diagonal
polarizability $\alpha_{1S}$ should also be modified by multiplying
the FSI factor $|1+2G_{\pi}(q^2)T^{I=0}_{\pi\pi,\pi\pi}(q^2)|^2$,
namely,
\begin{widetext}
\begin{equation}
\label{eq:wllfsi} d\Gamma(1^3S_1\to\pi^+\pi^-l^+l^-) =
\frac{(q^2-C)^2}{4b^2q_0^2}|1+2G_{\pi}(q^2)T^{I=0}_{\pi\pi,\pi\pi}(q^2)|^2
|\alpha_{1S}|^2\sqrt{1-\frac{4m_{\pi}^2}{q^2}} \sqrt{q_0^2-q^2}
\Gamma_{ee}(1^3S_1)dq^2dq_0.
\end{equation}
\end{widetext}
It was shown \cite{vo04} that by restricting the maximal value of
$q_0$ (or a lower cut-off on the invariant mass of the lepton pair)
to about 0.9 GeV, the higher intermediate state effect comes from
the $2S$ state only and can be evaluated through the experimental
data analysis by using Eq.(13) in Ref. \cite{vo04} without
additional parameters (except for an overall phase).

Now we are able to investigate the effects of the $\pi\pi$ FSI
corrected $|\alpha_{\psi'J/\psi}|$ on the scattering quantities in
the $J/\psi$-nucleon scattering. The $J/\psi$-nucleon total cross
section was measured as $\sigma_{J/\psi N}=3.8\pm0.8\pm0.5$ mb at
$\sqrt{s}\approx 5.7$ GeV \cite{ex77}. There are also some
theoretical calculations (see review of this subject in Ref.
\cite{st01}). For instance, Brodsky and Miller demonstrated the fact
that the two-gluons exchange in the scalar channel dominates the
elastic $J/\psi$-nucleon scattering and obtained $\sigma_{J/\psi
N}\approx 7$ mb at threshold \cite{sm97}. Based on the operator
product expansion and the trace anomaly, Teramond et al. found that
the total cross section at threshold is about 5 mb \cite{te98}. In
general, the range of predicted elastic $J/\psi+N \to J/\psi+N$
cross section at $p_{J/\psi}\approx 0$ GeV is $1.25\sim 5$ mb
\cite{st01}. Recently, in terms of the multipole expansion and the
low energy theorem, Sibirtsev and Voloshin showed the lower bound of
the total cross section at the threshold is about 17 mb \cite{sv05}.

It is known that the amplitude of the $J/\psi$-nucleon scattering at
low energies is proportional to the chromo-polarizability of
$J/\psi$, $\alpha_{J/\psi}$ \cite{sv05},
\begin{equation}
T_{J/\psi N} = \frac{16\pi^2}{9}(1+D)\alpha_{J/\psi}M_{J/\psi}m_N^2,
\end{equation}
where $D\geqslant1$ is a constant. Consequently, the scattering
length and the total cross section are proportional to
$\alpha_{J/\psi}$ and $\alpha_{J/\psi}^2$, respectively. The
in-medium mass shift of $J/\psi$ is equal to the real part of the
potential \cite{sv05,ha99}, and hence is also proportional to
$\alpha_{J/\psi}$. We present the estimated values of the scattering
length $a_{J/\psi}$, the cross section near the threshold
$\sigma_{J/\psi N}$ and the $J/\psi$ mass shift $\Delta M_{J/\psi}$
in Table \ref{tab:corr}. In the calculation, $\alpha_{J/\psi}=0.83$
GeV$^{-3}$ in the with $S$ wave $\pi\pi$ FSI case and
$\alpha_{J/\psi}=2$ GeV$^{-3}$ in the without $S$ wave $\pi\pi$ FSI
case \cite{sv05}, respectively .

\begin{table}[htb]
\caption{\label{tab:corr} Scattering length, cross sections and
$J/\psi$ mass shift ($\alpha_{J/\psi}=0.83$ GeV$^{-3}$ in the with
$S$ wave $\pi\pi$ FSI case and $\alpha_{J/\psi}=2$ GeV$^{-3}$ in the
without $S$ wave $\pi\pi$ FSI case \cite{sv05}).}
\begin{ruledtabular}
\begin{tabular}{lcc}
   & Without $\pi\pi$ FSI \cite{sv05} &  With $\pi\pi$ FSI \\ \hline%
$a_{J/\psi}$ (fm) & 0.37 & 0.15 \\%
$\sigma_{J/\psi N}$ (mb) & 17 & 2.9 \\%
$-\Delta M_{J/\psi}$ (MeV) &  21  & 8.7 \\%
\end{tabular}
\end{ruledtabular}
\end{table}

It should be noted that similar to those in Ref. \cite{sv05}, all of
the estimated results in Table \ref{tab:corr} are just the lower
bounds, because the lower bound values of $\alpha_{J/\psi}$ and $D$
are used. Thus, the estimated cross section of $J/\psi$-nucleon
scattering near threshold in Ref. \cite{sv05} is reduced from
$\gtrsim$ 17 mb to $\gtrsim$ 2.9 mb. Unlike that in Ref.
\cite{sv05}, this lower bound value agrees with the experimental
value of $\sigma_{J/\psi N}=3.8\pm0.8\pm0.5$ mb at $\sqrt{s}\approx
5.7$ GeV \cite{ex77} and compatible with the theoretical results in
Refs. \cite{sm97,te98,st01}.

In conclusion, the effect of the $S$ wave $\pi\pi$ FSI on
determining the chromo-polarizability is studied. The FSI corrected
formula for analyzing $J/\psi$ and $\Upsilon$ $\pi^+\pi^-$
transition data to get the chromo-polarizabilities of $J/\psi$ and
$\Upsilon$ is given. It is found that the effect of the $\pi\pi$ FSI
is quite sizeable, so that the values of chromo-polarizabilities
$|\alpha_{\psi'J/\psi}|$ and $|\alpha_{\Upsilon'\Upsilon}|$ can be
reduced to about 1/3 of those in the without $\pi\pi$ FSI case. As a
consequence, the scattering length and the in-medium mass shift of
$J/\psi$ are reduced to about 5/12 of the values given in Ref.
\cite{sv05} where $|\alpha_{\psi'J/\psi}|=2$ GeV$^{-3}$ was employed
\cite{vo04}, and the estimated lower bound of the total cross
section is reduced from 17 mb \cite{sv05} to 2.9 mb. We suggest that
in terms of Eq. (\ref{eq:wllfsi}), $\alpha_{J/\psi}$ should be
further measured in the $J/\psi\to\pi^+\pi^-l^+l^-$ decay at CLEO-c
and future BES-III, and $|\alpha_{\Upsilon'\Upsilon}|$ should be
further investigated in the $\Upsilon\to\pi^+\pi^-l^+l^-$ decay in B
factories. The precisely measured values would be very important in
studying the scattering of $J/\psi$ ($\Upsilon$) with light hadrons,
and hence in understanding the $J/\psi$ suppression in heavy-ion
collisions.

\begin{acknowledgments}
This work is partially supported by the NSFC grant Nos. 10475089,
10435080, CAS Knowledge Innovation Key-Project grant No. KJCX2SWN02
and Key Knowledge Innovation Project of IHEP, CAS (U529).
\end{acknowledgments}

\end{document}